# A Cubic Equation of State for Amyloid Plaque Formation

J. C. Phillips

Dept. of Physics and Astronomy, Rutgers University, Piscataway, N. J., 08854

Abstract

Protein function depends on both protein structure and amino acid (aa) sequence. Here we show that modular features of both structure and function can be quantified from the aa sequence alone for the amyloid 770 aa precursor protein A4. Both the new second order hydropathicity scale, based on evolutionary optimization (self-organized criticality), and the standard first order scale, based on complete protein (water-air) unfolding, yield thermodynamically significant features of amyloid function related to spinodal decomposition into modules. Spinodal crossings are marked by breaks in slope of algebraic functions of aa sequences. The new scale is associated with a lower effective temperature than the standard scale.

In his 1943 Dublin lectures "What is Life?", Erwin Schrodinger described an "aperiodic crystal" which could carry genetic information, a description credited by Francis Crick and James D. Watson with having inspired their discovery of DNA [1,2]. Schrodinger arrived at this picture from thermodynamic theories concerning protein stability and information content. Thus one can say that Schrodinger may have been the first theorist to conjecture that protein functionality could be usefully described thermodynamically. Here we take Schrodinger's reasoning one step further, by analyzing the aa sequence of the amyloid precursor, and showing that it contains strong thermodynamic features.

For a homologous group of proteins with a common fold, one can pass directly from (aa sequence) to functional relations, without attempting to elucidate all the intermediate structural stages, such as oligomer formation. A powerful tool in this analysis is an accurate hydropathic scale ψ(aa,1) that quantifies the small solvent-exposed areas associated with hydrophobic aa, reflecting their tendencies to be in the globular interior, and the much larger areas associated with hydrophilic aa, reflecting their tendencies to be on the surface. It is just here that Self-Organized Criticality (SOC) [3-5] comes to our aid, in the form of a remarkable discovery [6] that has gone almost unnoticed. One uses Voronoi partitioning to construct polyhedra centered on each aa in turn, and from these calculates the solvent accessible surface area A to a 2 Å spherical water molecule for each aa, averaged over a very large number (> 5000) of structures in the Protein Data Base.

If we average these A over complete protein structures, we assume that the water-driven areas are independent of the protein fold. Less restrictively, we may assume that the aa-specific areas depend on the length 2N + 1 of a modular protein segment centered on the aa, and plot A as a function of N. One then obtains [6,7] a stunning and very profound result: for 4 < N < 17, A ∝ $N^\delta$, which is power-law, self-similar scaling (SOC). The exponents δ define a universal SOC-based modular hydropathicity scale ψ(aa,1), which has turned out to be more useful in most cases than all older ψ scales, and much more accurate. Nor is that all: the lower modular limit of the power law range, N = 4, is probably related to the pitch of Pauling's α helix. The upper modular limit is typical of larger protein secondary structures (such as β sandwiches). The center of the modular range, N = 10, corresponds to a typical length of a cell membrane-spanning protein segment [8]. This scale accurately reflects the collective effects associated with SOC that are responsible for similarities in the dynamics of proteins and molecular glass-forming liquids [9].

The next step is to realize that because water-protein interactions are weak - compared to protein-protein backbone interactions, which are strong - their modular effects will be averaged over length scales W to obtain smoothed aa profiles ψ(aa,W). It is here that specific protein features enter the calculation. For flu proteins these features are either the spacing of relevant glycosidic (sugar) sites along the aa chain, or the length of molecules to which the viral protein is bound. These spacings are known from structural studies, and as a check on the method, they also are confirmed by conservation of ψ(aa,W) profile features under migration and vaccination pressures [7].

The precursor amyloid protein A4 of Alzheimer's disease (AD) (details available on Uniprot as P05067) has 770 aa. We begin by profiling the full 770 aa A4, which resembles a cell surface receptor, with a putative transmembrane region centered on 712. Here we expect the optimal value of W for describing membrane interactions to be W* = 21 (membrane thickness). In Fig. 1 we see that the amyloid fragments of A4 (672-711, Aβ 40, and 672-713, Aβ 42) are associated with a large, compound hydrophobic peak of ψ(aa,21) near the C terminal. The details of this compound 80 aa peak, and their relation to the structure of its β fragments [10] and plaque formation, will be discussed elsewhere.

The W-dependence of variances of ψ(aa,W) = $\mathcal{R}$(W), also called profile roughnesses, obtained from the SOC MZ [6] and the standard unfolding KD [11] scales [12] used in these calculations, provide insight into their thermodynamic significances as well as the stabilizing effects on A4 of the hydrophobic C terminal region. Broadly speaking, if the aa are uncorrelated, one should have -dln$\mathcal{R}$/dlnW ~ 1, according to the Central Limit Theorem [8], and for A4(1-670), this is valid within 20% for both scales, i.e., nearly indistinguishable $\mathcal{R}$(W). Specifically, for a random sequence one has $\mathcal{R}$W = const., an "equation of state" analogous to the ideal gas "equation of state", with W analogous to the volume V, and $\mathcal{R}$ analogous to the pressure p.

Plaque formation can be understood as a kind of precipitation that is occurring after fragmentation. The key step is the fragmentation one itself – what drives the full protein to produce the fragments that are then able to self-assemble into plaque? Surprisingly, the full A4 protein 1-770 profiles exhibit striking differences between the MZ and KD scales (Figs. 2a,2b) The MZ scale $\mathcal{R}$(W) exhibits linear softening in the 40 aa window between two breaks in slope at W = 19 and W = 61. This window suggests that the 80 aa compound hydrophobic peak stabilizes A4 by making it smoother and less exposed to conformational disruptions. The KD structure in $\mathcal{R}$(W) is much weaker.

The linear softening of $\mathcal{R}$(W) can be compared to a linear stability tie across the neck of a spinodal near a critical point, with the breaks in slope representing spinodal crossings. We can say that the $\mathcal{R}$(W) MZ and KD patterns of A4 resemble those of respectively lower and higher isothermals of the van der Waals (vdW) equation, which contains a spinodal region associated with its cubic character. The weakness of the KD structure suggests that its effective temperature T* is close to the critical temperature $T_c$. Spinodal decomposition typically produces composition waves with a wave length W [13]. Here the fragmentation wave begins at the C terminal with W ~ 80. Thus one can regard plaque formation as the end result of spinodal instability of the water-amyloid interface.

The Boolchand "reversibility window" seen in network glass melting transitions also has spinodal character, as its composition lies between soft, under- and rigid, over-constrained compositions. In the glass case abrupt changes in the nonreversible enthalpy of melting $\Delta H_{nr}$ occur at two spinodal crossings, with $\Delta H_{nr}$ dropping to nearly zero in the "reversibility window" between the crossings. The spinodal crossings are accompanied by vibrational thresholds with exponents in agreement with theory [14], as are the relaxation exponents of more homogeneous glasses, with lower fictive temperatures [15]. The nearly reversible glass network transition window is relevant to proteins, which can cycle reversibly through millions of transitions before failing. It could also be related to the A4 irreversible plaque-forming property, and even to Schrodinger's "aperiodic crystal".

Because the α helical pitch is < 4, the excluded molecular volume in this vdW model is $V_m = 7 = (2 \times 3 + 1)$ aa. Membranes both define cells and function as substrates for protein-protein interactions. Thus we define the critical volume as $W_c = 21$ aa, where the interaction length matches membrane thickness; these estimates satisfy the vdW relation $W_c = 3 V_m$. Our vdW amyloid equation is

$$(\mathcal{R} + 3\mathcal{R}_c(21/W)^2)(W - 7) = T^* \qquad (1)$$

with $\mathcal{R}_c = T_c^*/56$. The smallest epitopes for binding antibodies to the flu glycoprotein hemagglutinin are 7-mers [16]. The lower limit of the MZ modular range is 9, just above $V_m$. For specific proteins breaks in slope of $\mathcal{R}(W)$ may occur at other values of W, but the protein universe is still dominated by $V_m = 7$ and $W_c = 21$. One can regard the present connection of $V_m$ = 7 and $W_c$ = 21 to the modular picture as partial justification of the SOC properties discovered in [6].

The real value of the present theory lies in its ability to obtain, by simple algebraic means, using only aa sequences, thermodynamically significant aspects of plaque formation for the amyloid protein A4 (770 aa). Elsewhere we will exhibit similar thermodynamic aspects of the much smaller Aβ 40 aa protein fragments, as well as similar results for related amphiphilic calcitonin [17] and parathyroid hormone [18]. Detailed methods, such as molecular dynamics simulations (MDS) [19] and advanced modeling methods [20] often leave subtle functional trends unresolved, even for small proteins [12]. Thermodynamics, hierarchical scaling methods, unfolding and SOC work well for all sizes, here from 40 aa to 770 aa. Roughness and the thermodynamic differences between first-and second-order conformational changes are quantified in similar ways, independent of protein size.

The theoretical assumption underlying the present thermodynamic model is that the protein universe, in spite of its apparent diversity, has a larger degree of modular homogeneity [21] than one might have expected. Homogeneity facilitates synchronization of protein dynamics [22]. The universal features underlying modular protein functionality are helical stiffness, secondary lengths, and membrane thickness. The present thermodynamic model incorporates these features into hydropathic analysis, and shows that by doing so the MZ scale achieves a lower effective temperature [23] and higher resolution of functional and mutational trends. It may be useful for constructing protein phase diagrams that can be used to organize evolutionary data and engineer protein strains with desired properties [24].

*Postscript.* There is an extensive literature on amino acid pair interactions – corresponding to our W = 3, or N = 1 - and protein folding [25,26], which also has thermodynamic aspects, and semi-quantitatively reflects the importance of hydrophobic forces. The case N = 0 and W = 1 has been discussed bioinformatically [27]. We are grateful to Dr. D. C. Allan for comments and suggestions.

**87.14.em, 87.15.Cc, 87.15.kr, 87.15.bk**
**We construct a phase diagram for amyloid plaque formation, which determines the onset of Alzheimer's disease, using amino acid sequences alone. Our methods are general and can be applied to any protein family, without knowing the protein structure. Our methods transfer to protein networks many aspects of other compact networks, such as glass networks. We compare two scales, the standard scale (15,000 + citations), and a new scale based on self-similarity and self-organized criticality, and show that the new scale is superior. This biophysical paper has been submitted to PRE because it contains 8 PRL and 4 PRE references.**

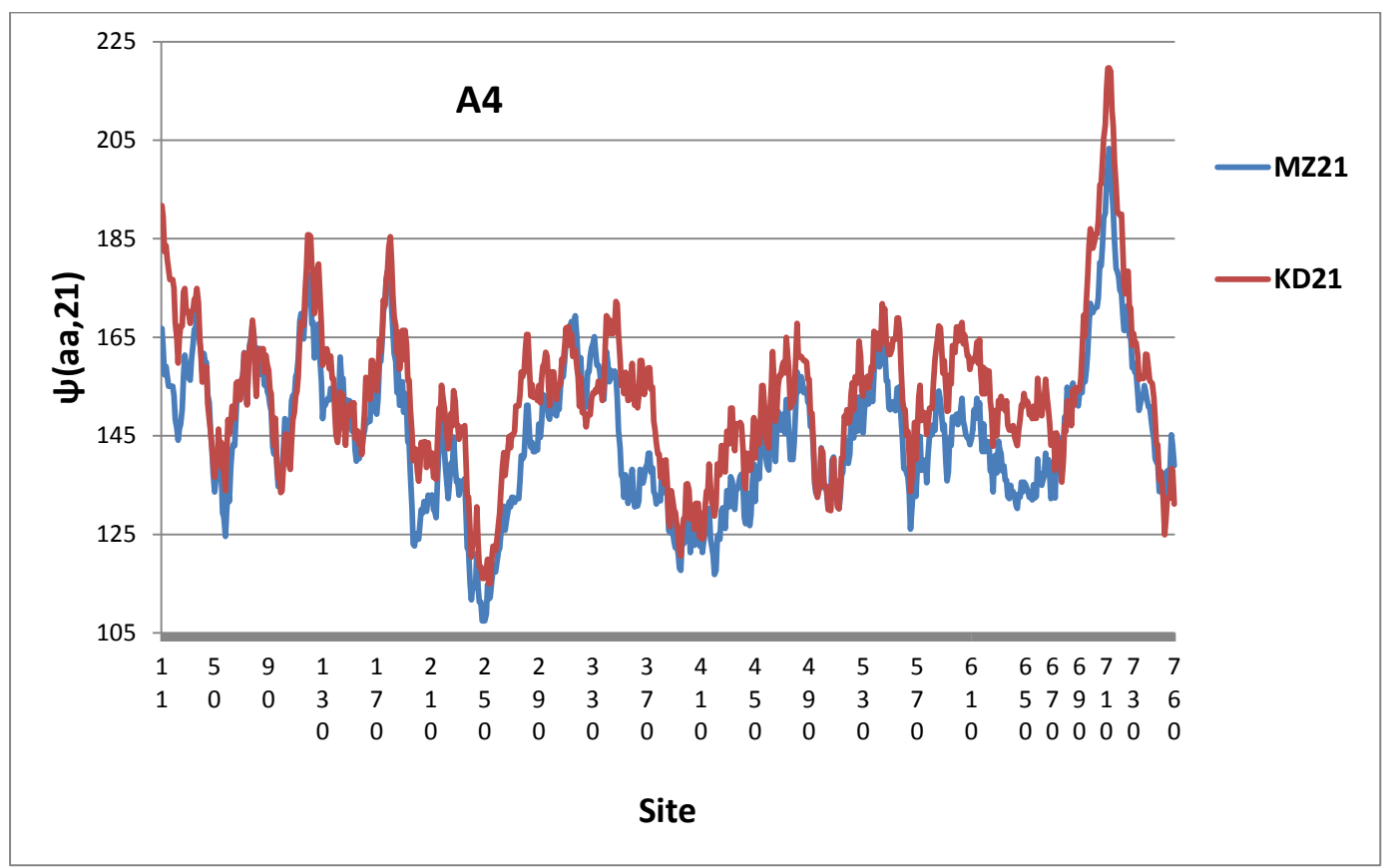

Fig. 1. Large-scale hydropathic profiles of ψ(aa,21) of the amyloid protein A4, using either the KD or MZ scales on the membrane modular length scale W = 21. The amyloid fragments occur in the 40 aa left half of the compound C-terminal hydrophobic peak. This 80 aa peak is strongly hydrophobic, and rises well above the profile of the rest of A4. It is also about four times as long

as a typical 20 aa transmembrane segment of a GPCR protein [8]. It splits in half to form the Aβ fragments, which then self-organize into plaque.

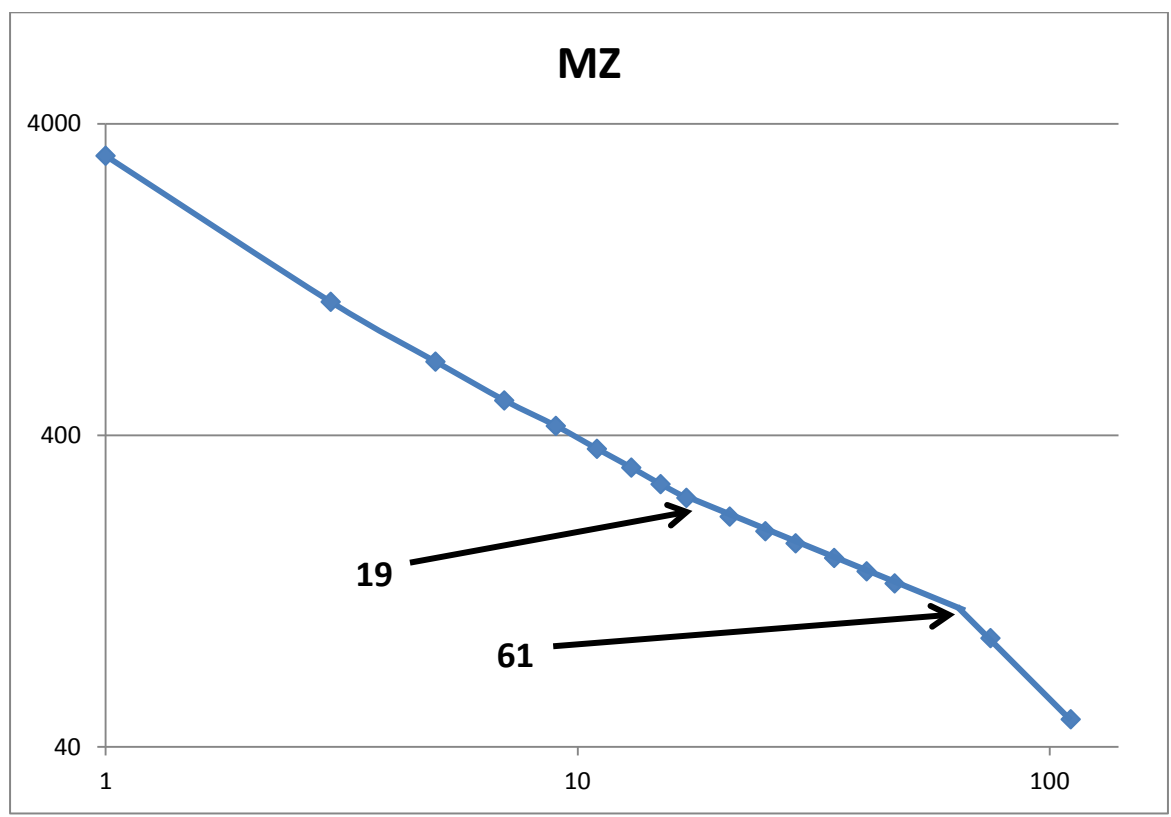

Fig. 2(a). A log-log plot of $\Re(W)$ for A4 (770 aa) obtained from the MZ scale reveals two breaks in slope, with a linear region between the breaks. In the text this linear region is assigned to a spinodal tie line.

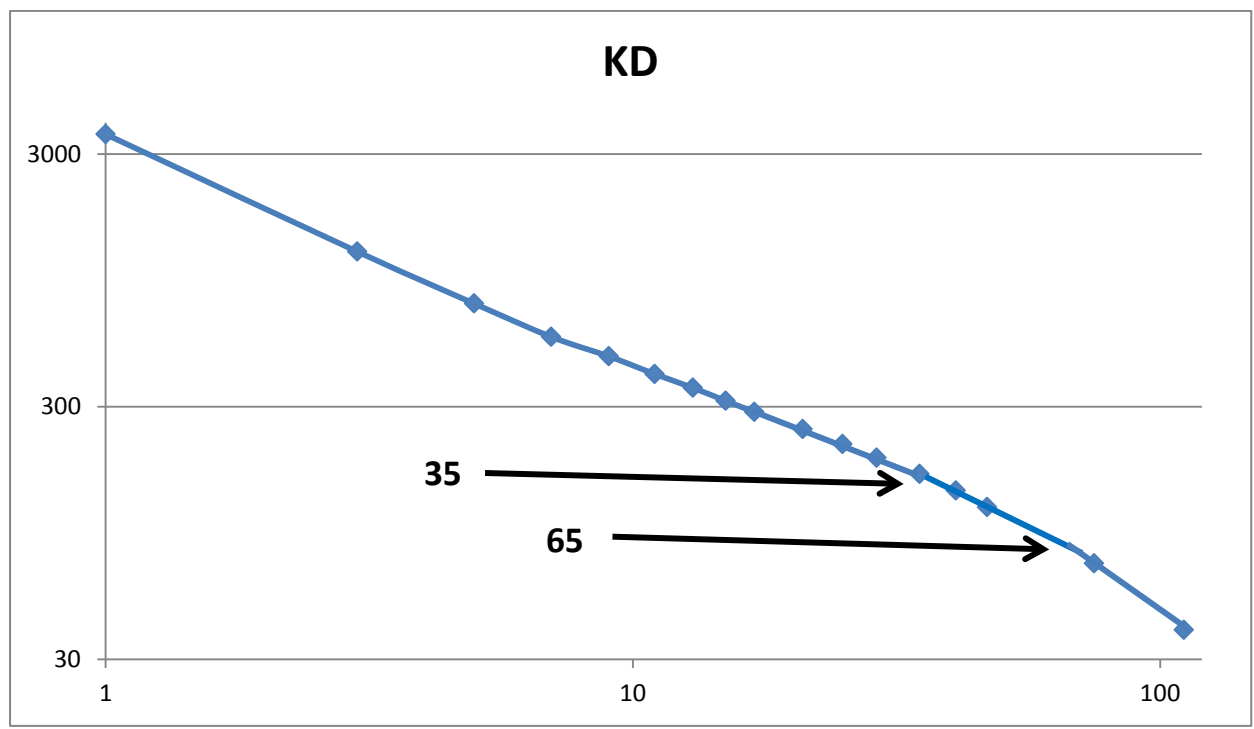

Fig. 2(b). A log-log plot of $\Re(W)$ for A4 (770 aa) obtained from the KD scale exhibits only weak breaks in slope. Accordingly the effective temperature T* is close to the critical temperature $T_c^*$, and spinodal effects are obscured by large thermal fluctuations.